\documentclass[twocolumn,trackchanges]{aastex701}

\newcommand\kms{km s$^{-1}$}
\newcommand\scm{s cm$^{-3}$}

\usepackage{amsmath}
\usepackage{multirow}

\shorttitle{{\it XRISM} Analysis of N103B}
\shortauthors{Holland-Ashford}
\accepted{February 5, 2026}
\submitjournal{ApJ}
\begin{document}

%\title{{\it XRISM} Analysis of Ejecta Kinematics and Thermal Properties in the Supernova Remnant N103B}
\title{{\it XRISM} Observation of the Supernova Remnant N103B: Velocity Structure and Thermal Properties}%Analysis of Ejecta Kinematics and Thermal Properties in the Supernova Remnant N103B}

\correspondingauthor{Tyler Holland-Ashford}
\author[orcid=0000-0002-7643-0504]{Tyler Holland-Ashford} 
\affiliation{Astrophysics Science Division, NASA Goddard Space Flight Center Greenbelt, MD 20771, USA}
\email[show]{tyler.e.holland-ashford@nasa.gov}

\author[orcid=0000-0003-2063-381X]{Brian J. Williams}
\affiliation{Astrophysics Science Division, NASA Goddard Space Flight Center Greenbelt, MD 20771, USA}
\email{brian.j.williams@nasa.gov}

\begin{abstract}

We present the first analysis of the X-ray Imaging and Spectroscopy Mission ({\it XRISM}) observation of the supernova remnant (SNR) N103B. We fit the X-ray spectrum taken with the Resolve microcalorimeter, which captured emission lines from the predominantly ejecta elements Si, S, Ar, Ca, Cr, Mn, and Fe. Notably, our fits require a previously unidentified high-temperature, highly-ionized, Fe-dominated plasma component with particularly high Cr and Mn abundances, matching a feature also present in the recent {\it XRISM} analysis of the SNR N132D. We find that all ejecta in N103B exhibits significant line broadening arising mostly from thermal Doppler broadening: increasing from $\sigma_{\rm th}\sim1700$ \kms\ for intermediate-mass element (IME: Si, S, Ar, and Ca) ejecta to $\sim$2800 \kms\ for Fe-rich ejecta. These velocities correspond to reverse shock velocities of $\sim$3500 and $\sim$5900 \kms, respectively, in the ejecta frame of rest. Finally, we find that the IMEs  are redshifted with a bulk velocity of $\sim$360 \kms\ while the Fe-dominated components are split: one redshifted at $\sim$1560 \kms\ and the other blueshifted at $\sim$1020 \kms. Our results provide further support for the double-ring structure of N103B as it expands into the bipolar winds of a non-degenerate companion and highlight the strength of high-resolution spectroscopic observations of SNRs.

\end{abstract}


\section{Introduction}
\label{sec:intro}

SNR 0509-68.7, also known as N103B, is a supernova remnant (SNR) in the Large Magellanic Cloud (LMC). It has a radius of $\sim$30\arcsec, corresponding to 7.2~pc at its distance of 50~kpc. It has a Sedov age estimate of 690 $\pm$ 20~yr \citep{ghavamian17}, consistent with the light echo estimate of 400--880~yr \citep{rest05} and the expansion measurement age of $\lesssim$860~yr \citep{williams18}. N103B is thought to be the remnant of a Type Ia explosion \citep{hughes95,lewis03,lopez11,yang13,yamaguchi14} with a relatively massive companion based on the star formation history around this SNR \citep{badenes09} and the evidence of a dense circumstellar medium (CSM) the SNR is interacting with (e.g.,  \citealt{williams14,someya14,li17}). Based on analysis of the CSM abundances, the companion star is suggested to have been a main-sequence star \citep{blair20}.

The western half of the SNR is the brightest and contains emission from the SNR's interaction with dense CSM \citep{williams14,yamaguchi21}. Emission from this half of the remnant is redshifted \citep{li17,yamaguchi21}, indicating that it is on the far side of the SNR. In the southeast, the SNR appears to be interacting with a giant molecular cloud with H$_2$ density even higher than that in the west \citep{sano18}. Emission from ejecta material in the east appears to be blueshifted \citep{yamaguchi21}. 

The similarities between N103B and Kepler's SNR have led astronomers to suggest similar origin scenarios for the two SNRs. 
Both are Type Ia SNRs exhibiting evidence of strong CSM interaction, have similarly high post-shock densities \citep{williams14}, and their morphologies exhibit a strong dipolar asymmetry despite most Type Ia SNRs being more spherically symmetric \citep{lopez11}. 

The X-ray spectrum of N103B shows strong lines from elements typically seen in Type Ia SNRs (e.g., Si, S, Ar, Ca, and Fe; \citealt{someya14}), with O and Mg also present but whose distribution better matches with CSM clumps rather than that of other SNR-synthesized ejecta materials \citep{guest22}. \cite{yamaguchi21} performed spatial and spectral analysis of the {\it Chandra} X-ray emission from N103B and discovered a double-ring structure present in the intermediate-mass element (IME; Si, S, Ar, Ca) ejecta which suggests that the SNR is expanding into an hourglass-shape cavity resulting from strong bipolar winds from a single-degenerate progenitor binary system. 

In this paper, we analyze the Resolve microcalorimeter X-ray spectrum from N103B. We focus on analyzing the kinematic properties---line broadening and bulk Doppler shift---of both the bulk ejecta and individual ejecta elements, information which can only be measured with the $\lesssim$5 eV spectral resolution of Resolve. In Section~\ref{sec:observations}, we present the data used and our process for reprocessing it. In Section~\ref{sec:anal}, we explain our methods of analyzing the data and present our results. In Section~\ref{sec:disc}, we derive reverse shock velocities and ion temperatures and discuss how our results compare to previous studies on N103B in the literature. 

\begin{figure*}[ht!]
\begin{center}
\includegraphics[width=0.8\textwidth]{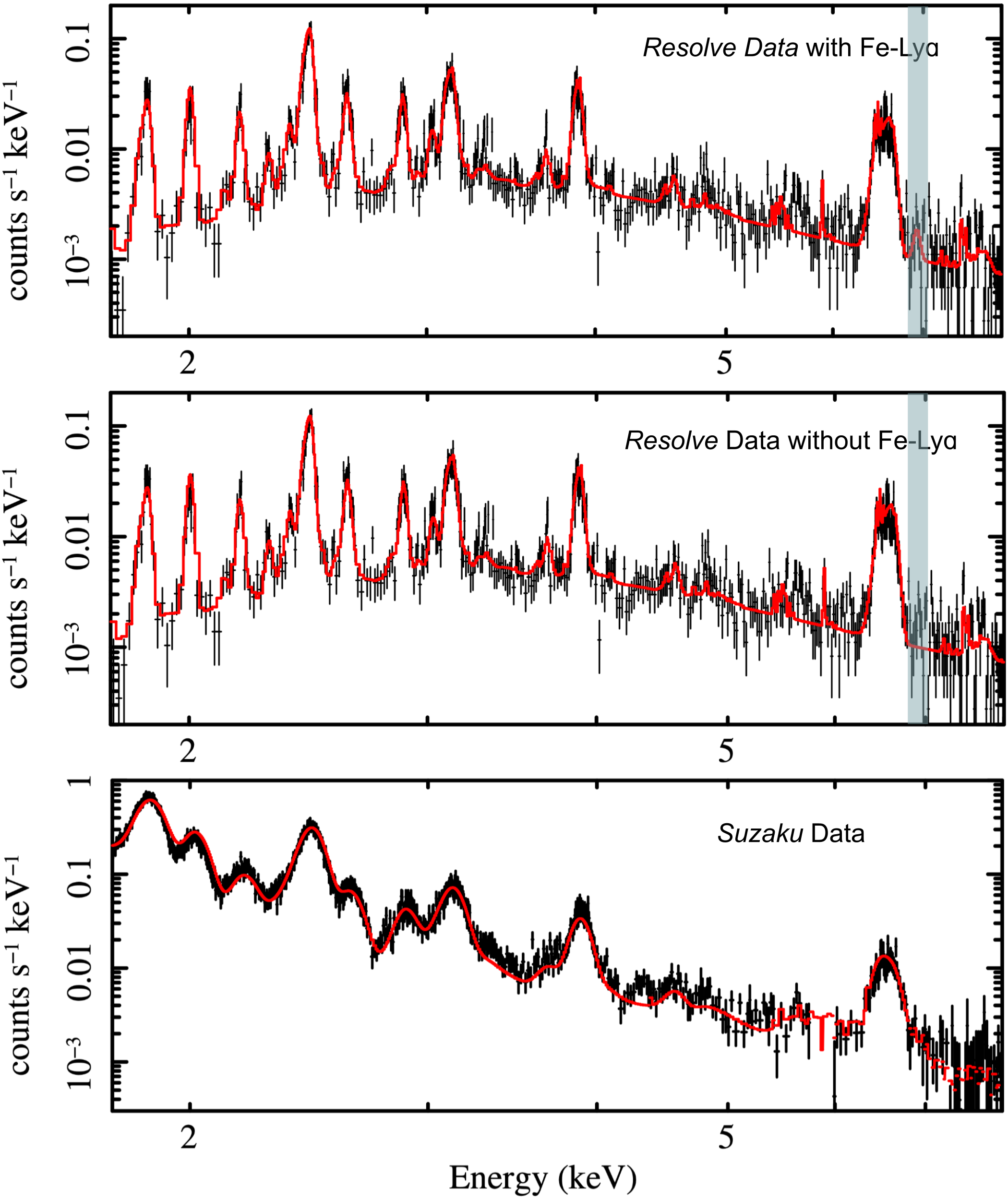}
\caption{\footnotesize{ The 1.75--8.0~keV spectrum and best fit model of the SNR N103B, showing (Top): Resolve data to which the model is fit, (Middle): the same Resolve data, but without the dedicated 10~keV component to capture the Fe Ly$\alpha$ emission at 6.95~keV (the grey highlighted region), and (Bottom): {\it Suzaku} data overlaid with a model fit to Resolve data and including an added systematic uncertainty of 0.1 to reflect effective area uncertainties. %We note that the ionization timescale of the Resolve-fit model does not quite match the Si emission (1.75--2.1~keV) as measured by {\it Suzaku}.
In addition to the optimal binning of \cite{kaastra16} used during fitting, the data is binned by an additional factor of 4 for plotting purposes only.}
\label{fig:FullSpec_rsl}}
\end{center}
\end{figure*}

\section{Observations and Data Reduction}
\label{sec:observations}

We analyzed the 190-ks {\it XRISM} observation of N103B beginning on 2024 Sep 30 (ObsID 201104010), reprocessing the data following the steps in the {\it XRISM} ABC Guide\footnote{\url{https://heasarc.gsfc.nasa.gov/docs/xrism/analysis/abc_guide/xrism_abc.html}}, using CALDB files released on 2025 Mar 15, and using HEASoft version 6.35.1. We included only Grade 0 (i.e., high-resolution primary: Hp) events from all pixels except 12 and 27 in our analysis, resulting in a Hp branching ratio of $>$99\%. We created a large-sized redistribution matrix file (RMF)\footnote{Using an extra-large sized RMF resulted in similar preliminary results but analysis took significantly longer.} and used a {\it Chandra} image of N103B in the 0.5--8.0~keV bandpass to generate ancillary response file (ARF). We accounted for the contribution of non X-ray background (NXB) spectrum according to the steps outlined in Section~6.6 of the {\it XRISM} ABC Guide: fitting a model to the NXB spectrum generated with the \texttt{rslnxbgen} tool using the location of N103B and times corresponding to the dates of our observations. When fitting the observed spectra of N103B, we included this NXB model as an additive frozen component so that our best-fit spectral parameters captured only emission from the SNR itself.
Finally, we used the optimal binning option of the \texttt{ftgrouppha} task in FTOOLS to group the counts in our resolve spectrum \citep{kaastra16} prior to fitting. 

Additionally, to obtain constraints on the SNR emission below $\sim$2~keV, we analyzed the 234-ks observation of N103B taken with {\it Suzaku} in June 2016 (ObsID 804039010). We followed the steps in the {\it Suzaku} Data Reduction Guide\footnote{\url{https://heasarc.gsfc.nasa.gov/docs/suzaku/analysis/abc/}}, using the {\it Suzaku} calibration database (CALDB) files released in 2018 Oct 23, the HEADAS software version 6.35.1, and the reprocessing tool \texttt{aepipeline}. Given the high signal-to-noise of the data, we only used the spectrum extracted from the XIS0 front-illuminated detector.

\begin{figure*}[ht!]
\begin{center}
\includegraphics[width=0.8\textwidth]{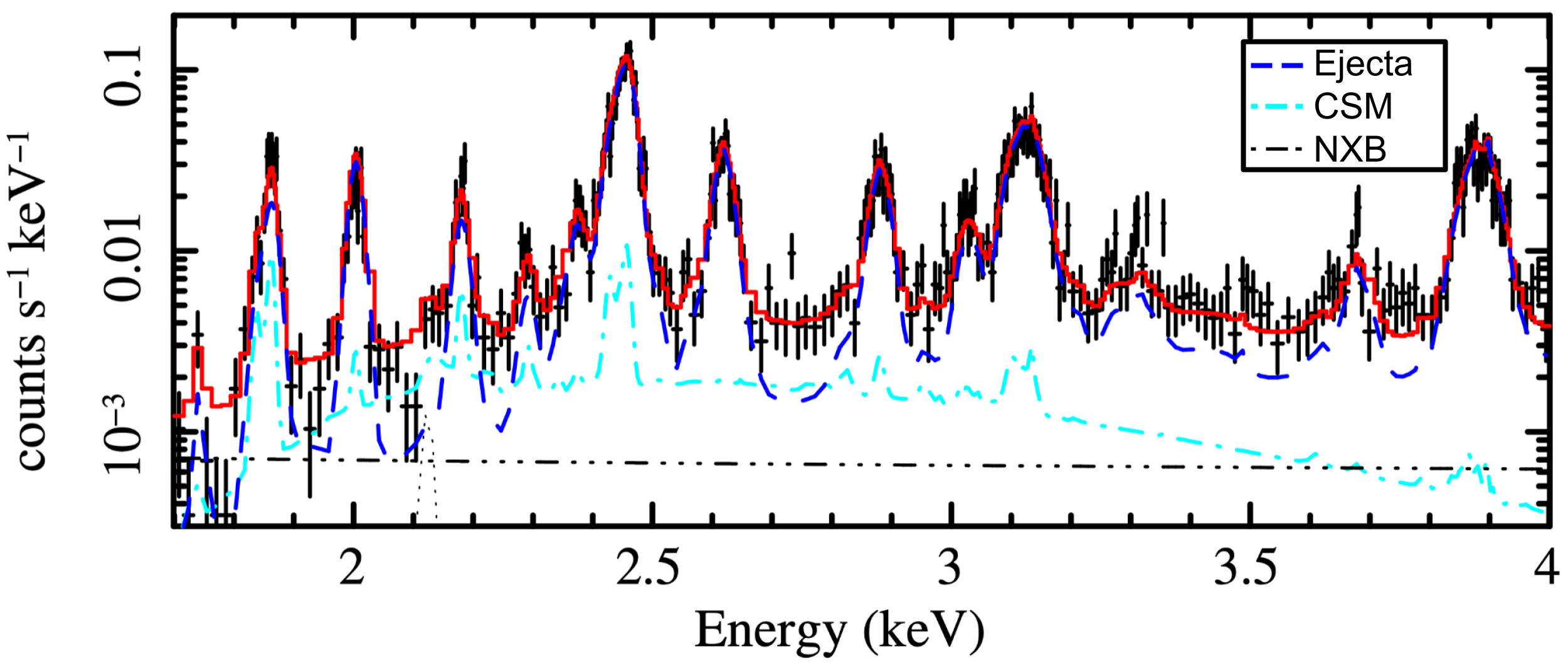} \\
\caption{\footnotesize{The 1.7--4.0~keV Resolve spectra of N103B, fit with non-equilibrium ionization models including variable components for line width, Doppler shift, ionization timescale, electron temperature, and ejecta abundances. The blue dashed line captures emission from shocked ejecta elements: mainly Si, S, Ar, and Ca. The cyan dash-dotted line captures the swept-up CSM/ISM emission found by fitting {\it Suzaku} data, and the black lines are the contributions from the non X-ray background. In addition to the optimal binning of \cite{kaastra16} used for fitting, the data is binned by an additional factor of 4 for plotting purposes only. 
}
\label{fig:IME_fit}}
\end{center}
\end{figure*}

\section{Data Analysis}
\label{sec:anal}

We used \texttt{XSPEC} version 12.15.0 and AtomDB version 3.1.3 to fit the spectrum of N103B. Our best-fit model comprised two absorption components (Galactic: \texttt{TBabs} and LMC: \texttt{TBvarabs}) and four velocity-broadened non-equilibrium ionization (NEI) collisional plasma components (\texttt{bvnei} or \texttt{bvvnei}). Using fewer plasma components produced worse fits and using more didn't improve the fit. These four NEI components included one low-temperature component to capture emission from shocked circumstellar and interstellar material (CSM \& ISM), one moderate temperature component to capture emission from IME-dominated ejecta (``ej1''), and two high-temperature components to capture emission from Fe-rich ejecta (``ej2'' and ``ej3''). We note that the secondary Fe-rich component was not detected in any previous papers analyzing data from CCD-resolution telescopes, but was necessary to capture emission lines at 6.95~keV corresponding to the Fe-Ly$\alpha$ emission line in the Resolve spectrum. Figure~\ref{fig:FullSpec_rsl} shows the 1.75--8.0~keV Resolve X-ray data overplotted with a model including and excluding this Fe Ly-$\alpha$ component.

To obtain constraints on the components that dominate at lower energies (i.e., the hydrogen column density, swept-up material, and elements less massive than Si), we initially fit this model to the 0.6--8.0~keV bandpass of the {\it Suzaku} data to obtain a rough best-fit and then froze all parameters of those lower-energy components when fitting the Resolve spectrum. The ejecta abundances of the 0.5~keV component capturing swept-up material were frozen to those found in \cite{someya14}. For this fitting process, we added a systematic uncertainty of 10\% to capture the effects of the effective area calibration uncertainties \citep{marshall21, ha23, ha25}. The bottom panel in Figure~\ref{fig:FullSpec_rsl} shows the {\it Suzaku} data overplotted with our best-fit Resolve model.

The Resolve spectrum shown in Figure~\ref{fig:FullSpec_rsl} displays distinct emission line features from Si, S, Ar, Ca, Cr, Mn, and Fe. The broadening of the lines---much greater than Resolve's spectral resolution of $\sim$5~eV, but much smaller than {\it Suzaku's} spectral resolution of $\gtrsim$50~eV---reflects the inherent broadening of the emitting material. We dedicate the majority of this paper to analyzing the properties of emission lines in this spectrum: their Doppler shifts (reflecting bulk ejecta velocity), velocity broadenings (reflecting a mix of thermal Doppler broadening and multiple ejecta clumps moving at different velocities), and the relative line strengths between different emission lines (reflecting plasma temperature and ionization timescale).

In addition to the NEI plasma fitting analysis described in the following sections, we also fit the observed emission lines with Gaussians. We describe this process and the results of it in Appendix~\ref{appen:gaussians} and note that we obtained similar results as our NEI analysis.

\subsection{IME Ejecta}
\label{sub:IME_NEI}
To obtain constraints on intermediate-mass elements, we focused on the 1.7--4.0~keV bandpass that contained emission from Si, S, Ar, \& Ca. We fit this region with our model, fixing the Fe abundance to the value found by the 0.6--8.0~keV spectral fit ([Fe]/[Fe]$_\odot \approx 0.37\times 10^5$) and freezing all parameters in the Fe-dominated components (``ej2'' and ``ej3''). The central portion of Table~\ref{tab:NEI_IME_Fitting} in the Appendix shows our best-fit values for this model, and Figure~\ref{fig:IME_fit} shows the best-fit spectrum.

We found that the IME-dominated plasma is best fit with an electron temperature around 2.2~keV, an ionization timescale of $\sim$10$^{11}$ \scm, a Doppler shift of $\sim$1.2$\times10^{-3}$ (corresponding to a bulk velocity of $\sim$360 \kms), and a velocity broadening of $\sim$1730 \kms. 

We then fit smaller bandpasses of the spectrum to identify the thermal properties of individual elements. We performed this analysis for the 1.6--2.33~keV (Si emission lines), 2.39--3.0~keV (S emission lines), and 2.93--4.7~keV (Ar \& Ca emission lines) bandpasses. We combined the Ar \& Ca emission lines into a single fit due to the lower signal of that bandpass. The best-fit parameters of our plasma models are reported in the bottom three sections of Table~\ref{tab:NEI_IME_Fitting} in the Appendix and visually displayed in Figure~\ref{fig:IME_individ_fit}. 

Our measured IME velocities---bulk Doppler shift and velocity broadening---for IMEs are presented in Table~\ref{tab:Measured_Velocities}. 
The uncertainties reported here and throughout this paper are 90\% confidence limits found via the Xspec \texttt{error} command. 

All of these values are consistent to within 2$\sigma$ of each other and the values found from the 1.7--4.0~keV fit, except for the velocity broadening of the Si emission which is $>$2$\sigma$ below the broadening of any other element. Given the low effective area of Resolve below 2~keV and the fact that there is the most contamination from the swept-up CSM at energies $<$2~keV, it is not surprising that the Si emission would differ the most. However, contamination from CSM emission should broaden the observed Si lines: the opposite of what we measured.

\begin{figure*}[h!]
\begin{center}
\includegraphics[width=0.8\textwidth]{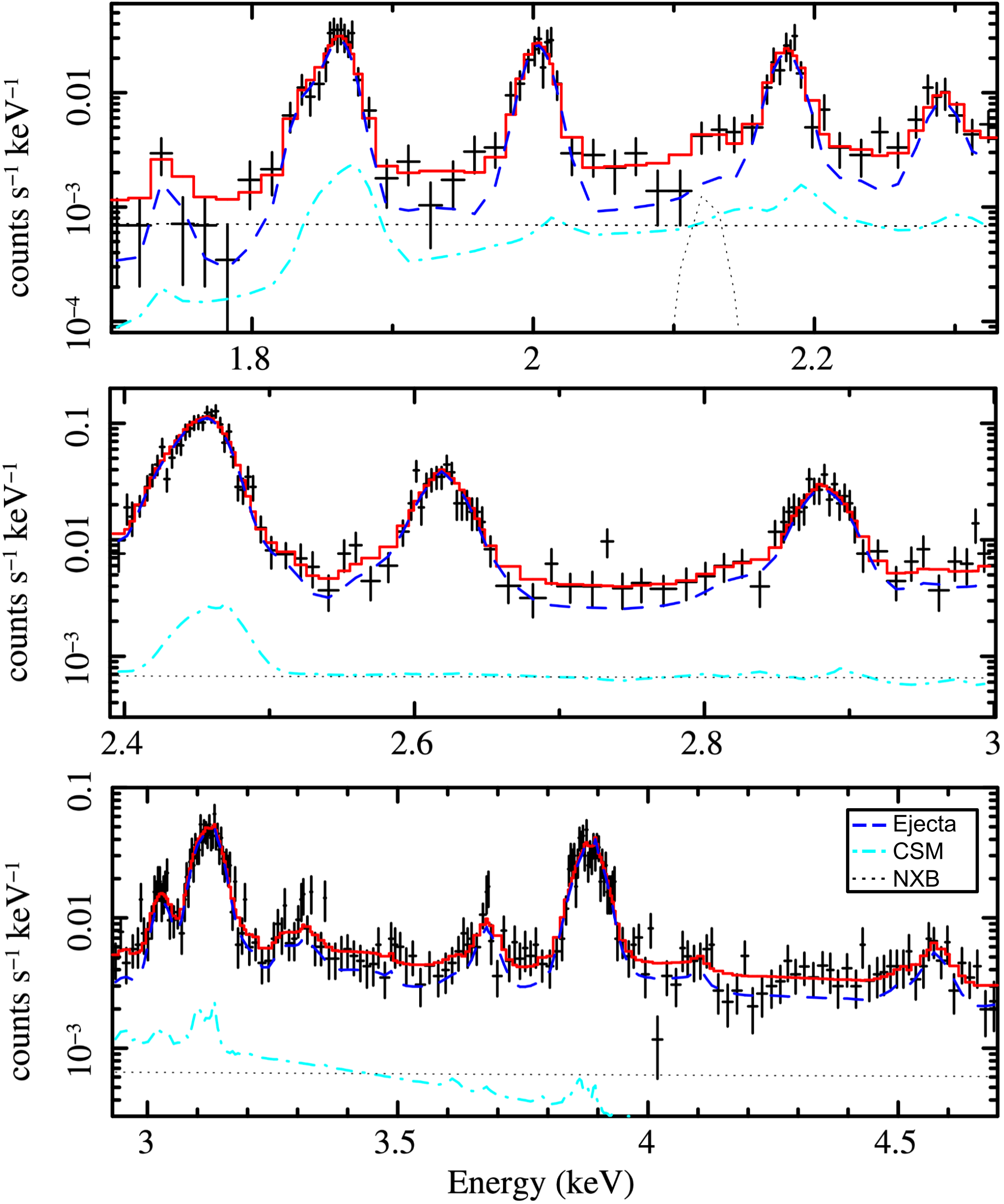}  \\
\caption{\footnotesize{ {\it Top}: The 1.7--2.33~keV (Si emission lines), {\it Middle}: The 2.39--3.0~keV (S emission lines), and {\it Bottom}: 2.93--4.7~keV (Ar \& Ca emission lines) Resolve spectra of N103B, fit with non-equilibrium ionization models. The blue line captures emission from shocked ejecta emission, the cyan dash-dotted line captures the swept-up CSM/ISM emission, and the black dotted lines are the contributions from the non X-ray background. In addition to the optimal binning of \cite{kaastra16} used for fitting, the data is binned by an additional factor of 4 for plotting purposes only. 
}
\label{fig:IME_individ_fit}}
\end{center}
\end{figure*}

\subsection{IGE NEI Fitting}
\label{sub:IGE_NEI}
To obtain precise plasma properties of the iron group elements (IGEs: Cr, Mn, Fe, \& Ni), we fit the 3.5--7.38~keV bandpass that included Ca emission lines and continuum emission dominated by ej1. This allowed us to let the Fe abundance, normalization, ionization timescale, line broadening, and Doppler shift from ej1 vary freely---necessary to capture the full profile of the Fe-K complex in N103B---while ensuring that our best-fit model remained consistent with the flux at lower energies. When we tried to fit this model to a bandpass comprising only IGEs (e.g., 5.0--7.4~keV), we were unable to obtain constraints on many plasma properties
---particularly ej1's Fe emission. With a wider bandpass (e.g., 1.7--7.4~keV), the fit would settle in a local minimum that prioritized matching the lower energy, higher signal-to-noise emission, resulting in inaccurately measured IGE properties.

\begin{deluxetable*}{l|c|c}
\tablewidth{0pt}
\tablecaption{Measured Ejecta Velocities\label{tab:Measured_Velocities}}
\tablehead{
\colhead{Element} & \colhead{Doppler Shift} & \colhead{Velocity Broadening}  \\  \colhead{} & \colhead{(\kms)} &\colhead{(\kms)} 
}
\startdata
All IMEs & 360$^{+93}_{-102}$ &  1730$^{+80}_{-90}$  \\ 
Si &  391 $\pm$ 183 &  1460$^{+180}_{-170}$  \\
S &   282 $\pm$ 141 & 1740$^{+120}_{-110}$  \\ 
Ar\tablenotemark{a} &   \multirow{2}{*}{450$^{+207}_{-231}$} & \multirow{2}{*}{1760$^{+180}_{-160}$}  \\
Ca\tablenotemark{a} &   &  \\ 
Fe (ej2) &  -1030$^{+430}_{-640}$ & 2730$^{+680}_{-550}$  \\
Fe (ej3) &  1570$^{+310}_{-790}$ & 2900 (fixed)  \\\hline
\enddata
\tablenotetext{a}{Doppler shifts and velocity broadening are tied together for Ar \& Ca in our NEI model.}
\tablecomments{\footnotesize{All uncertainties are 90\% confidence intervals and all parameters are rounded to contain three significant digits.}}
\vspace{-6mm}
\end{deluxetable*}

\begin{figure*}[ht!]
\begin{center}
\includegraphics[width=0.75\textwidth]{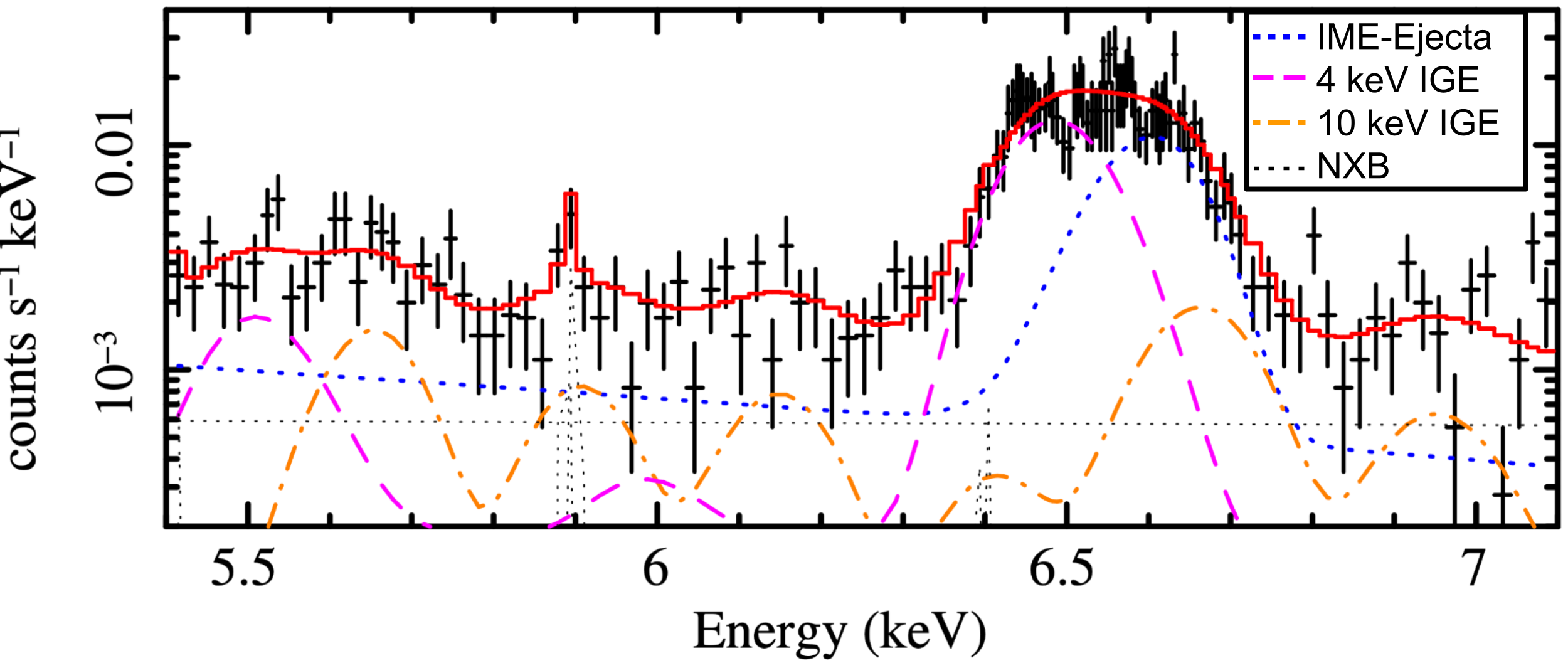} \\
\caption{\footnotesize{The 5.4--7.1~keV Resolve spectra of N103B fit with non-equilibrium ionization models to the 3.5--7.38~keV bandpass. The blue dashed line captures emission from the colder IME-dominated plasma component. The magenta dashed line captures emission from the lowly-ionized IGE-dominated component. The orange dash-dotted line captures emission from the high temperature, highly-ionized IGE-only component. The black dotted lines are the contributions from the non X-ray background. In addition to the optimal binning of \cite{kaastra16} used for fitting, the data is binned by an additional factor of 4 for plotting purposes only. 
}
\label{fig:IGE_fit_LyalphaCrMn}}
\end{center}
\end{figure*}

As noted previously, we had to add a second, more highly-ionized IGE-dominated component (``ej3'')  to capture the Fe-Ly$\alpha$ emission line at $\sim$6.95~keV. Following the method of \cite{xrism24}---who found a similar component to be necessary for fitting the LMC SNR N132D---we included a 10~keV \texttt{bvvrnei} component with [Fe] fixed to 10$^5$. We allowed the ionization timescale, redshift, normalization, and abundances of Cr and Mn to vary freely in this component, but had to fix the velocity broadening to 2900 \kms\ (approximately that of the other Fe-dominated component) to allow the fit to converge. The spectral fit for the IGE-bandpass is presented in Figure~\ref{fig:IGE_fit_LyalphaCrMn} and the best-fit parameters can be found in Table~\ref{tab:NEI_IGE_Fitting} in the Appendix.

Our measured IGE velocities are presented in Table~\ref{tab:Measured_Velocities}. We note that emission from the lower-temperature, lower-ionization IGE component (``ej2'') is blueshifted, while emission from the higher-temperature IGE component (''ej3'') is redshifted. Additionally, we note that this latter component had best-fit [Cr/Fe] and [Mn/Fe] abundance ratios of $>$20 times that of the cooler IGE-dominated component, indicating its importance if one wants to obtain an accurate estimate of IGE ejecta mass ratios.

\section{Discussion}
\label{sec:disc}
From the above spectral analysis, we obtained estimates of ejecta properties including ionization timescales, bulk ejecta velocities, and velocities from Doppler broadening of ejecta. In this section we discuss the implications of our findings, how they relate to previous studies of N103B, and present the thermal broadening, reverse shock velocity, and ion temperatures that can be derived from our best-fit parameters.

\subsection{Ejecta Kinematics}
We see an increase in both velocity broadening and bulk velocity with increasing ejecta mass, going from $\sigma_{\rm v, Si}\approx1460$ \kms\ to $\sigma_{\rm v, Ar}\approx1760$ \kms, and v$_{\rm S}\approx280$ \kms\ to v$_{\rm Ar}\approx450$ \kms.
This trend is further strengthened when considering the IGE results, with $\sigma_{\rm v, Fe}\approx2800$ \kms\ and v$_{\rm Fe} \approx$ 1000-1500 \kms: larger than those of IMEs by $\sim$1000 \kms.

The finding that more massive elements exhibit larger bulk velocities is
inconsistent with simulations that typically find lighter elements such as Si and S to have more extreme velocities than IGEs (e.g., \citealt{maeda10, seitenzahl13,wongwathanarat17}). However, observations of SNRs commonly find Fe-rich knots to be protruding beyond intermediate mass ejecta or even the forward shock (e.g., \citealt{fang18,orlando25}). A potential resolution to this apparent dilemma is that the supernova explosion was highly asymmetric, and thus the more massive elements formed in the innermost layers are subject to strongest asymmetric explosive forces (e.g., as in Kepler's SNR, \citealt{sato20}). Our results support this scenario, as we found that the IME and the highly-ionized IGE are redshifted with velocities of $\sim$400--1500 \kms, while the low-ionized IGE is blueshifted with a velocity of $\sim$1000 \kms.

We can compare this finding to what has been observed in Cas~A, where the ejecta can be divided into redshifted and blueshifted components, with the bulk of blueshifted ejecta moving nearly opposite to the bulk of the redshifted ejecta in the 2D plane of the sky \citep{grefenstette17,picquenot21}. Analysis with {\it XRISM} confirmed these findings and revealed that the IME and Fe Doppler shifts were roughly constant as a function of radial distance, providing support for incomplete shocked shell structures rather than bipolar-like expansion \citep{bamba25}. Additionally, \cite{bamba25} found that, while the redshifted IME and Fe Doppler velocities were similar, the blueshifted Fe ejecta had a velocity $\sim$1000 \kms\ larger than the blueshifted IMEs---a similar velocity difference as we found in N103B. Finally, the blueshifted ejecta is characterized by a higher percentage of heavy elements and the neutron star in Cas~A appears to be moving directly opposite this ejecta \citep{ha20,picquenot21}. 

\subsubsection{Spatially-Resolved Spectral Analysis}
While our Resolve (and Xtend) data does not have the resolution to perform detailed spatially-resolved analysis of this LMC SNR (R$_{\rm N103B}$ = 30'' $<$ Resolve's 1.3' HPD), there have been a previous studies analyzing {\it Chandra} spectra of N103B. 

\cite{yamaguchi21} used the Generalized Morphological Components Analysis method (GMCA: \citealt{picquenot19}) to analyze the spatial and spectral properties of N103B, identifying CSM emission, IME ejecta, and Fe ejecta. Similar to the results of \cite{guest22}---who also analyzed {\it Chandra} data---\cite{yamaguchi21} found that IME ejecta dominated around the rims of N103B and Fe-rich ejecta dominated the central regions. They interpreted their results as evidence for a double-ring structure (similar to SN~1987A) resulting from strong bipolar progenitor winds. This type of progenitor system is also suggested for Kepler's SNR \citep{burkey13,chiotellis20}, but while Kepler's SNR is viewed nearly edge-on, N103B is viewed at an angle of $\sim$45$^{\circ}$ so that the bipolar shells are only partially overlapping in the projected view (see Figure~11 of \citealt{yamaguchi21}). 

In our analysis of the Resolve spectrum, we found that most of the IME ejecta emission is redshifted. This matches the findings of \cite{yamaguchi21} and supports their suggestion of a double-ring structure. The western ejecta on the far side of the SNR is running into denser material and, as emission scales with density squared, is thus brighter. It is possible that this bright emission is masking a lower-flux, blueshifted IME-dominated component.

Our IGE results are also consistent with the proposed model of \cite{yamaguchi21}. We find two IGE-rich components, with high velocities in opposite directions, supporting the existence of a bipolar explosion where the ejecta is mainly expanding into two opposite lobes. The fact that we do not find low-velocity Fe might be due to a combination of the existence of a dense torus that is absorbing emission from the slowest-moving IGE ejecta or that the slowest-moving IGE ejecta is still unshocked. We also found that the redshifted IGE has a higher ionization timescale than the blueshifted IGE: consistent with the redshifted ejecta expanding into a higher density ambient medium (e.g., the dense CSM emission) and reaching  equilibrium more rapidly.

Finally, \cite{court25} explored interactions between Type Ia SNR ejecta and post-common envelope cocoon models with bipolar structures. They found that Kepler's SNR can be explained by an explosion into an ambient medium with uniform (albeit high) density, whereas N103B requires dense CSM interactions that significantly boost the ionization of shocked ejecta elements. This matches the asymmetric picture presented by \cite{yamaguchi21}.

\subsection{Ion Temperatures \& Shock Kinematics}
\label{subsec:disc_iontemps}
Given that \cite{yamaguchi21} found Si ejecta to be present along the rims of the SNR, we can use the relation found in \cite{xrism24} to derive the blast wave velocity (V$_{\rm bw}$) of N103B as a function of the observed velocity broadening $ \sigma_{\rm v}$.
\begin{equation}
   \sigma_{\rm v} = \frac{3}{8} V_{\rm bw}
\end{equation}
We use our observed Si velocity broadening of 1460 \kms\ to derive a blast wave velocity of V$_{\rm bw} \approx 3900$ \kms. This is consistent with the estimates of 2860--5450 (average of 4170) \kms\ for different locations of N103B's blast wave as measured by \cite{williams18}. We note that this calculation assumes that most of the Si broadening is either due to CSM being heated by the blast wave or due to reverse shock heating of ejecta that happened soon after the reverse shock was formed. 

We cannot use the same equation for the higher-mass elements as they are farther interior in the SNR and thus heated by the reverse shock many years after its formation. For these interior ejecta elements, our measured velocity broadening is a combination of thermal Doppler broadening from reverse shock heating ($\sigma_{\rm th}$) and variations in bulk velocity of different ejecta clumps ($\sigma_{\rm kin}$).
\begin{equation}
   \sigma_{\rm v} = \sqrt{\sigma_{\rm th}^2 + \sigma_{\rm kin}^2}
\end{equation}
We measured the total velocity broadening $\sigma_{\rm v}$, but we need the thermal broadening $\sigma_{\rm th}$ to derive the reverse shock velocity, which will in turn allow us to derive the ion temperatures of the shock-heated ejecta:
\begin{equation}
   \sigma_{\rm th} = \sqrt{\frac{k_bT_i}{m_{\rm i}}}
\end{equation}
where k$_b$ is the Boltzmann constant, and T$_i$ is ion temperature, and m$_i$ is ion mass.

To determine $\sigma_{\rm th}$, we use the equations presented in \cite{xrism24} that relate velocity broadening and reverse shock properties:
\begin{equation}
   \sigma_{\rm v} = \sqrt{\frac{13}{64}\left(\frac{R_{\rm rs}}{t}\right)^2 - \frac{18}{64}\frac{R_{\rm rs}}{t}V_{\rm rs}+\frac{21}{64}V_{\rm rs}^2}
\end{equation}
\begin{equation}
    v_{\rm u,sh} = \frac{R_{\rm rs}}{t} - V_{\rm rs} = \frac{4}{\sqrt{3}}\sigma_{\rm th}
\end{equation}
R$_{\rm rs}$ is the reverse shock radius (and thus R$_{\rm rs}$/t approximates the undeccelerated velocity of ejecta just hitting the reverse shock), V$_{\rm rs}$ is the reverse shock velocity in the observer frame, and $v_{\rm u,sh}$ is the upstream ejecta velocity in the shock-rest frame (i.e., the reverse shock velocity in the ejecta-rest frame).

We simplify these two equations by solving for R$_{\rm rs}$/t in Equation 5 and substituting that into Equation 4, 
\begin{equation}
    \sigma_{\rm v} = \sqrt{ \frac{13}{12}\sigma_{\rm th}^2 + \frac{1}{2\sqrt{3}}\sigma_{\rm th}V_{\rm rs} +\frac{1}{4}V_{\rm rs}^2}
\end{equation}
leaving us with a single equation with two unknown variables ($\sigma_{\rm th}$ and V$_{\rm rs}$). To solve this equation, we make assumptions that allow us to obtain minimum and maximum estimates of the thermal broadening.

The maximal thermal broadening occurs when there is no kinetic broadening, such that $\sigma_{\rm th} = \sigma_{\rm v}$. The minimum thermal broadening occurs when the reverse shock is moving the most slowly in the ejecta rest frame ($v_{\rm u,sh}$), corresponding to moving quickly outward in the observer reference frame (V$_{\rm rs}$). Reverse shocks in supernova remnants initially move outward in the observer frame, and then decelerate until eventually moving inward \citep{truelove99,micelotta16}. Although SNR N103B's age of $\sim$880~yr is relatively young, it is encountering dense ambient material which would result in a more rapid deceleration of the reverse shock. As the observer-rest frame reverse shock velocity in Cas~A (t$_{\rm age}\sim350$~yr) is found to be $\sim$1000--3500 \kms\ \citep{vink22,fesen25}, we assume a conservative upper limit of V$_{\rm rs} <$ 1500 \kms\ for N103B due to its higher age and ambient density. 

Taken together, we have the following limits
\begin{equation}
    \sigma_{\rm th}(V_{\rm rs} = 1500 \ {\rm km\ s}^{-1}) \leq \sigma_{\rm th} \leq \sigma_{\rm v} 
\end{equation}
which we substitute into Equation 6 to bound the possible values of the thermal broadening $\sigma_{\rm th}$. We take these limits as a 90\% confidence interval which we propagate with the uncertainties obtained from Xspec fitting to obtain total uncertainties. Finally, we plug our derived values for $\sigma_{\rm th}$ into Equations 5 and 3 to obtain estimates of the ejecta-rest frame reverse shock velocity $v_{\rm u,sh}$ and ion temperature kT$_{\rm i}$, respectively.

\begin{deluxetable*}{l|c|c|c}
\tablewidth{0pt}
\tablecaption{Derived Velocities and Ion Temperatures\label{tab:Derived_Velocities}}
\tablehead{
\colhead{Element} & \colhead{Thermal Broadening\tablenotemark{a}} &  \colhead{Ejecta-Rest Frame} & \colhead{Ion Temperature\tablenotemark{c}} \\
\colhead{} & \colhead{$\sigma_{\rm th}$ (\kms)}  & \colhead{Reverse Shock Velocity\tablenotemark{b}} & \colhead{kT$_{\rm i}$ (MeV)} \\
\colhead{} & \colhead{}  & \colhead{$v_{\rm u,sh}$ (\kms)} & \colhead{}  
}
\startdata
All IMEs &  1520 $\pm$ 170 &  3520 $\pm$ 390 & 0.68 $\pm$ 0.15 $(\frac{m_{\rm i}}{m_{\rm Si}})$ \\ 
Si &   1240$^{+250}_{-240}$ &  2860$^{+570}_{-540}$ & 0.45$^{+0.19}_{-0.16}$ \\
S &    1530$^{+200}_{-180}$ &  3520$^{+450}_{-420}$ & 0.77$^{+0.22}_{-0.17}$  \\ 
Ar\tablenotemark{a} &    \multirow{2}{*}{1560$^{+230}_{-220}$} &  \multirow{2}{*}{3590$^{+540}_{-510}$} & \multirow{2}{*}{1.00$^{+0.32}_{-0.28}$}  \\
Ca\tablenotemark{a} &    & \\ 
Fe (ej2) &   2530$^{+690}_{-570}$ &  5850$^{+1570}_{-1310}$ &  3.71$^{+2.28}_{-1.48}$ \\
Fe (ej3) &  2700$^{+710}_{-720}$\tablenotemark{d} &  6230$^{+1640}_{-1650}$\tablenotemark{d} & 4.22$^{+2.51}_{-1.94}$\tablenotemark{d}  \\\hline
\enddata
\tablenotetext{a}{Derived by plugging the measured total velocity broadening ($\sigma_{\rm v}$; see Table~\ref{tab:Measured_Velocities}) and the constraints regarding minimum and maximum thermal broadening (see Section~\ref{subsec:disc_iontemps}) to Equation 6. }
\tablenotetext{b}{Derived by plugging the thermal broadening values into Equation 5.}
\tablenotetext{c}{Derived by plugging the thermal broadening values into Equation 3.}
\tablenotetext{d}{Frozen during fitting, but here we assume a velocity broadening 90\% CI of $\pm$ 700 \kms, chosen to be slightly larger than that of the other Fe-dominated component. }
\tablecomments{\footnotesize{All uncertainties are 90\% confidence intervals.
}}
\end{deluxetable*}

We report our derived thermal broadening velocities, ejecta-rest frame reverse shock velocities, and ejecta ion temperatures in Table~\ref{tab:Derived_Velocities}. For bulk IMEs, the unknown reverse shock velocity contributed 86\% of the total uncertainty and we find that $\gtrsim$75\% of the total velocity broadening measured arises from thermal broadening due to a reverse shock velocity of $\gtrsim$3000 \kms\ in the frame of the expanding ejecta. For IGEs, the unknown reverse shock velocity contributed $\sim$22\% of the total uncertainty and we estimate that $\gtrsim$90\% of the total velocity broadening arises from thermal broadening due to a reverse shock velocity of v$_{\rm u,sh}\gtrsim$5400 \kms.
 
We note that the higher-temperature IGE component (ej3) had higher [Cr/Fe] and [Mn/Fe] abundances than the lower-temperature IGE-dominated component and was more highly-ionized than either of our other plasma components. The higher ionization timescale requires that this ejecta is either more dense, interacting with a denser ambient medium, or was shocked less recently. The latter two scenarios, along with the high bulk Doppler velocity and higher Cr and Mn abundances, suggest that this plasma component reflects fast-moving ejecta knots from a deeper layer of the explosion that have penetrated shells of lower-mass ejecta, as has been seen in Fe ejecta breaking through lighter-element rings in Cas~A (e.g., \citealt{orlando25}) or overtaking even the forward shock in Tycho's SNR (e.g., \citealt{fang18}).

\section{Conclusion}
We have analyzed the first {\it XRISM} observation of the supernova remnant N103B, studying the 1.75--8.0~keV spectrum taken by the Resolve microcalorimeter. We measured the Doppler-broadened velocities, bulk Doppler shifts, and ionization timescales of both bulk ejecta emission and that of individual elements. We found that the intermediate-mass ejecta elements (Si, S, Ar, \& Ca) were redshifted with velocities of $\sim$400 \kms\ and exhibited Doppler-broadened velocities that increased with ejecta mass: from 1460 \kms\ for Si to 1760 \kms\ for Ar \& Ca. The Fe emission in N103B requires three plasma components: one from the lower-temperature IME-dominated component and two from IGE-dominated components at higher temperatures. The first IGE-dominated component is lowly-ionized (n$_e$t $\approx 7.7 \times 10^9$ \scm) and blueshifted. The second IGE-dominated component is highly-ionized (n$_e$t $\approx 5.5 \times 10^{11}$ \scm), redshifted, and contains a much larger amount of Cr and Mn. Both exhibit bulk velocities much larger than that of the IMEs, and all of the line broadenings seem to be largely due to thermal Doppler broadening resulting from reverse shock heating.

We found that $\gtrsim$80\% of the line broadening measured is likely thermal broadening due to the ejecta being heated by the reverse shock, with the remaining $\lesssim$20\% a result of line-of-sight velocity variations of ejecta clumps. However, we note that the equations used to derive these estimates assume spherical symmetry. In the frame of the outward-moving ejecta, the velocity of the reverse shock is $\sim$3520 for IME and $\sim$5850 for the faster-moving IGE. Our measured velocity broadenings are consistent with reverse shock velocities in the frame of the observer of $-$1500 to $+$1500 \kms. 

The kinematics of N103B match with the morphology presented in \cite{yamaguchi21}: that of a explosion with a double ring structure due to expanding into strong bipolar progenitor winds and viewed at a $\sim$45$^\circ$ angle. The redshifted material is expanding into dense CSM, the blueshifted material is expanding into a molecular cloud, and a torus from the progenitor accretion disk might be blocking the innermost ejecta emission.

With {\it XRISM}'s gate valve currently closed, our analysis is limited to $\gtrsim$1.7~keV. If the gate valve were to open in the future, we could analyze the emission below 1.7~keV containing emission from elements lighter than Si as well as Fe-L emission. This latter emission is particularly important, as in preliminary analysis where we attempted to jointly fit CCD-resolution (e.g., {\it Suzaku}, {\it Chandra}, or Xtend) and Resolve X-ray data, we found it difficult to obtain acceptable spectral fits that simultaneously fit both the Fe-L and Fe-K emission. Each region seemed to require different ionization timescales and/or Fe abundance. It is unclear whether this issue is a result of using improper or insufficiently complex spectral models or whether it is an issue with inaccurate atomic data. In subsequent work, we plan to more deeply investigate this issue in order to determine if additional lab experiments to more robustly measure the emission lines of iron are necessary.

The Advanced X-ray Imaging Satellite ({\it AXIS}), a proposed NASA Probe Explorer mission, would have larger effective area (especially at energies $\lesssim$1.7~keV) and much higher spatial resolution than {\it XRISM}. It would allow us to characterize the lower energy emission from N103B as well as perform spatially-resolved spectroscopy to try to identify regions coincident with red- or blue-shifted ejecta. In general, future X-ray missions that have both high spatial and spectral resolution (e.g., NewAthena, Lynx) are vital to understand the full 3D kinematics of supernova remnants, particularly those of more distant SNRs in the Magellanic Clouds. Alternatively, joint analysis of {\it Chandra} and {\it XRISM} data could prove revolutionary for obtaining images that contain high spectral and spatial resolution. One such joint likelihood deconvolution (Jolideco; \citealt{donath24}) has been successfully shown to combine the spatial information from different telescopes' observations expanding it to also account for spectral resolution would greatly expand the science yield of {\it XRISM}.

%% The "ht!" tells LaTeX to put the figure "here" first, at the "top" next
%% and to override the normal way of calculating a float position.
%% The asterisk after "figure" tells the compiler to span multiple columns
%% if a two column style is selected.

%\url{http://journals.aas.org/authors/aastex.html}.

%\aastex\ v6 introduced five new table features that were designed to make
%table construction easier and the resulting display better for AAS Journal
%authors. The items are:

% \begin{enumerate}
% \item Declaring math mode in specific columns,
% \item Column decimal alignment, 
% \item Automatic column header numbering,
% \item Hiding columns, and
% \item Splitting wide tables into two or three parts.
% \end{enumerate}

% Full details on how to create each of these special table types are given in the guidelines at \url{http://journals.aas.org/authors/aastex.html}.

%% Please use the acknowledgment and contribution environments. This will 
%% be anonomyized when the "anonymous" style option is used. 
\begin{acknowledgments}
Tyler Holland-Ashford’s research was supported by an appointment to the NASA Postdoctoral Program at the NASA Goddard Space Flight Center, administered by Oak Ridge Associated Universities under contract with NASA. We thank the referee for their helpful comments that improved the clarity of the paper.
\end{acknowledgments}

\facilities{{\it XRISM}}

%This research made use of the analysis software: HEASoft (6.31, \small http://heasarc.gsfc.nasa.gov/ftools, \normalsize ascl:1408.004), XSPEC (v12.13.0; ascl:9910.005 \citealt{arnaud96} ), AtomDB (v3.1.3 \citealt{smith01,foster12}), \normalsize and CALDB (\small https://heasarc.gsfc.nasa.gov/FTP/caldb, \normalsize XRISM 20250315).

\software{HEASoft v6.35.1 \& CALDB \citep{2014ascl.soft08004N},  
          XSPEC v12.15.0 \citep{arnaud96}, 
          AtomDB v3.1.3 \citep{smith01,foster12}}


\appendix

\section{Best-fit NEI Parameters}
Here we report the full best-fit parameters of our Xspec plasma models along with the 90\% confidence intervals. Table~\ref{tab:NEI_IME_Fitting} below contains the plasma components used to fit the 1.7--4.0~keV bandpass of the spectrum (capturing emission from IMEs), and Table~\ref{tab:NEI_IGE_Fitting} contains the plasma components used to fit the 3.5--7.38~keV bandpass of the spectrum (capturing emission from IGEs).

\section{Gaussian Fitting}
\label{appen:gaussians}

In addition to NEI plasma model fitting, we initially fit the observed emission lines with Gaussians. For this process, we created an Xspec model with a thermal \texttt{bremss} component to capture continuum emission and Gaussian components to capture known emission lines. We froze the Gaussians' centroid energies to the energies of known atomic emission lines taken from AtomDB and applied the Xspec model component \texttt{vashift} to account for Doppler shifts due to bulk ejecta velocity. Due to the large line broadening present in the spectrum of N103B we were unable to leave all the parameters of each Gaussian independent. Thus, for all emission lines from the same element, we applied a single Doppler shift and tied the Gaussian line widths to scale linearly with centroid energy. 

Our results for fitting Gaussians to the spectrum are displayed in Figure~\ref{fig:Gaussian_Fits}, with best-fit parameter values and their 90\% confidence intervals reported in Table~\ref{tab:Gaussian_Values}. For tractability, we split the bandpasses into three: Si \& S (1.75--3.0~keV), Ar \& Ca (3.0--4.65~keV), and iron group elements (IGEs: Cr, Mn, \& Fe; 5.1--7.38~keV). When multiple lines were indistinguishable from a single broadened line (e.g., the Si-He$\alpha$ intercombination doublet, Si-Ly$\alpha_1$ and Si-Ly$\alpha_2$), we set the fixed centroid energy of the Gaussian to the weighted average of the centroid energies of the combined lines. Overall, our Gaussian fitting returned results consistent with our IME fitting. 

\subsection{Caveats of Gaussian Fitting}

We note that emission from swept-up material is strongest at energies $<$2~keV and thus contaminates ejecta emission from lighter ejecta elements. Thus, the true velocity broadening for Si ejecta is likely lower than our reported values, and our reported uncertainties are underestimates. For Ca, the lower signal-to-noise, fewer amount of lines prominent lines, and overlap with emission lines from other elements all combined to make it likely that certain Gaussians were incorrectly capturing the emission from other emission lines, resulting in artificial broadening and line shifts. Thus, the true errors of our best-fit parameters for Ca are likely larger than the reported uncertainties.

\begin{figure*}[ht!]
\begin{center}
\includegraphics[width=0.8\textwidth]{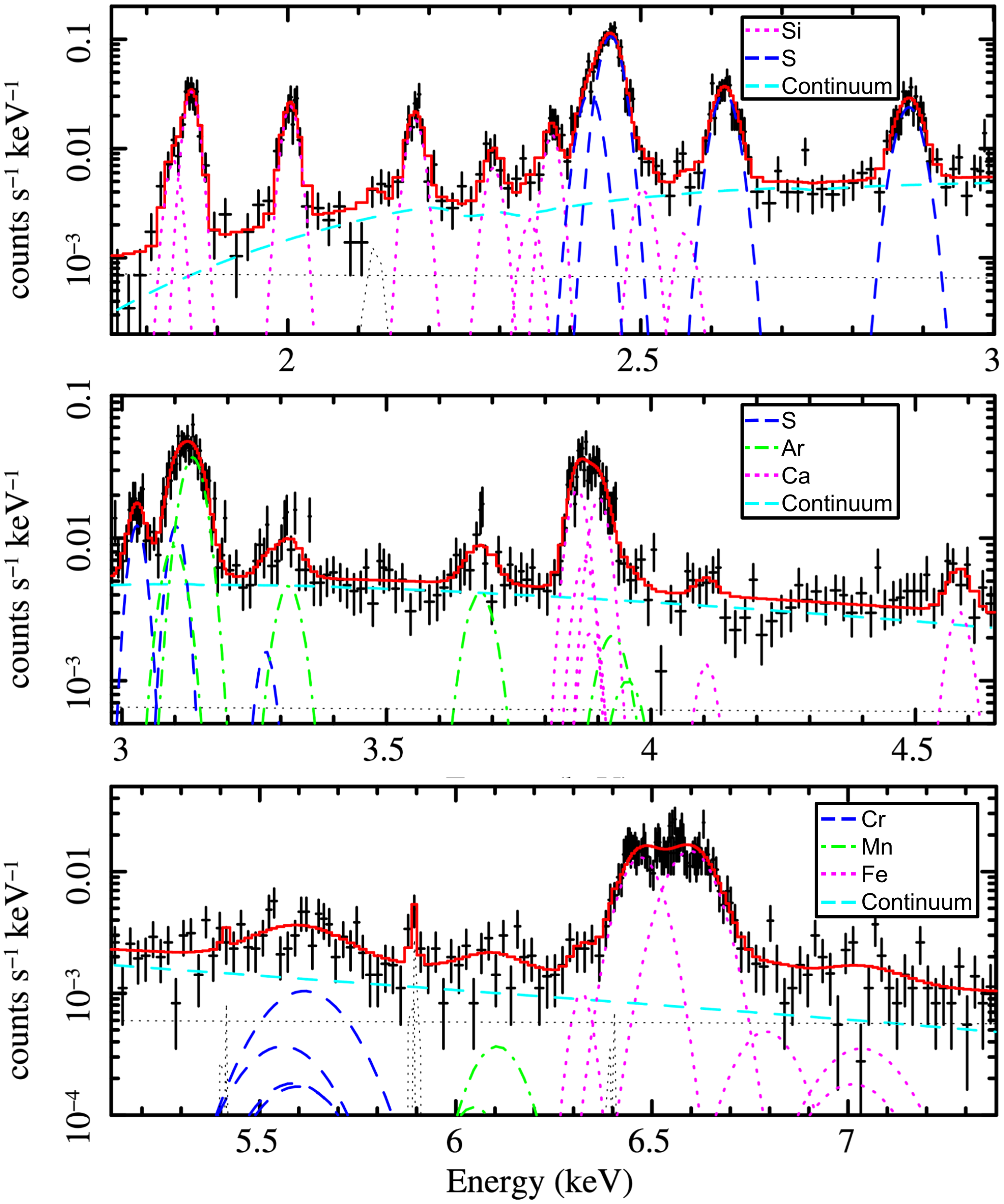} \\
\caption{\footnotesize{ {\it Top}: The 1.75--3.0~keV Resolve spectrum of N103B, fit with Gaussians to capture emission lines from silicon (magenta dotted) and sulfur (blue dashed).  {\it Middle}: The 3.0--4.65~keV Resolve spectrum of N103B, fit with Gaussian lines to capture emission from sulfur (blue dashed), argon (green doted-dashed), and calcium (magenta dotted).  {\it Bottom}: The 5.1--7.38~keV Resolve spectrum of N103B, fit with Gaussians to capture emission lines from chromium (blue dashed), manganese (green dash-dotted), and iron (magenta dotted). The cyan dashed line is a bremsstrahlung model fit to the continuum, and the black dotted lines are the contributions from the non X-ray background. The Cr and Mn lines have normalizations set to match the emissivity ratios at 6.845~keV. In addition to the optimal binning of \cite{kaastra16} used for fitting, the data is binned by an additional factor of 4 for plotting purposes only. }
\label{fig:Gaussian_Fits}}
\end{center}
\end{figure*}

Our ability to obtain accurate Gaussian fits to the Fe-K complex at 6.4--6.7~keV was hampered by the large broadening of the Fe emission lines present in the spectrum. As shown by the bottom panel of Figure~\ref{fig:Gaussian_Fits}, our best-fit models would settle on using only a few Gaussians to capture the Fe-K complex from 6.4--6.7~keV. Given the vast number of emission lines in this region and the unknown effects of velocity shift(s) and velocity broadening(s), we could not place any constraints on the kinematics and ion temperature of Fe using this Gaussian fitting method. Similarly, we could not constrain the properties of Cr or Mn emission lines, even when tying the normalizations of the four most prominent emission lines---the forbidden, intercombination doublet, and recombination---from each element together. Finally, the emission from Ni Ni at $\sim$7.75~keV was too faint to be detectable above the noise and is not included in our results or analysis.

The issues presented in the preceding paragraphs highlight some of the issues with using Gaussians to fit the X-ray spectra of supernova remnants. If there is significant line blending due thermal broadening or the existence of multiple ejecta components at different velocities along the line-of-sight, it becomes difficult, if not impossible, to characterize individual features. This regime is where plasma models excel, as the best-fit plasma temperature and ionization timescale constrain the relative normalizations between emission lines from different ions. However, we still include details on our Gaussian fitting process for thoroughness sake, as an independent check on the best-fit parameters obtained via NEI model fitting, and as a repository of X-ray emission lines.

\begin{deluxetable}{lcc}
\tablewidth{0pt}
\tablecaption{NEI Fitting to IMEs \label{tab:NEI_IME_Fitting}}
\tablehead{
\colhead{Parameter} & \colhead{Best-Fit Value (90\% CI)} 
}
\startdata
$N_{\rm{H, MW}}$ (10$^{22}$ cm$^{-2}$)  & 0.062 (fixed) \\
$N_{\rm{H, LMC}}$ (10$^{22}$ cm$^{-2}$)  & 0.3 (fixed) \\ \hline \hline
Fitting to Bandpass containing IME ejecta (1.7--4.0 keV) \\
Swept-Up CSM/ISM: \texttt{bvnei} \\
kT$_e$ (keV) & 0.546 (fixed) \\
Ionization Timescale ($10^{13}$ cm$^{-3}$ s)  & 5 (fixed) \\
Redshift ($10^{-3}$)  & 1.5 $^{+2.7}_{-1.8}$ \\
Velocity Broadening (\kms)  & 719$^{+615}_{-394}$\\
Normalization	($10^{-2}$ cm$^{-2}$)   & 1.82 $\pm$ 0.72 \\ \hline
Lower-temperature Ejecta Component: \texttt{bvnei} \\
$kT_{\rm e}$ (keV)    & 2.21$^{+0.54}_{-0.49}$ \\
Abundance (10${^5}$ solar) O--Mg & 0.3 (fixed) \\
%Ne  & 1 (fixed) \\
%Mg  & 3 (fixed)  \\
Si  & 1 (fixed) \\
S  & 1.09 $\pm$ 0.12\\
Ar  & 1.06 $\pm$ 0.16\\
Ca  & 1.54 $\pm$ 0.24\\
Fe  & 0.37 (fixed) \\
Ionization Timescale($10^{11}$ cm$^{-3}$ s) & 1.10$^{+0.46}_{-0.19}$\\
Redshift ($10^{-3}$)  & 1.20$^{+0.31}_{-0.34}$ \\
Velocity Broadening (\kms)  & 1733$^{+80}_{-86}$\\
Normalization	(10$^{-7}$ cm$^{-5}$)   & 3.08 $\pm$ 0.19 \\ \hline \hline
%%%%%%%%%%%%%%%%%%%%%%%%%%%%%%%%%%%%%%%%%%%%%%%%%%%%%%%%%
Fitting to Si Bandpass (1.6--2.3~keV) \\
$kT_{\rm e}$ (keV)    & 1.62$^{+1.32}_{-0.58}$ \\
Ionization Timescale($10^{11}$ cm$^{-3}$ s) & 1.11$^{+2.04}_{-0.52}$\\
Redshift ($10^{-3}$)  & 1.38 $\pm$ 0.61 \\
Velocity Broadening (\kms)  & 1460$^{+184}_{-168}$\\
Normalization	(10$^{-7}$ cm$^{-5}$)   & 3.40$^{+2.34}_{-1.12}$ \\ \hline \hline
%%%%%%%%%%%%%%%%%%%%%%%%%%%%%%%%%%%%%%%%%%%%%%%%%%%%%%%%%
Fitting to S Bandpass (2.39--3.0~keV)  \\
$kT_{\rm e}$ (keV)    & 1.69$^{+0.47}_{-0.31}$ \\
Ionization Timescale($10^{11}$ cm$^{-3}$ s) & 1.84$^{+0.73}_{-0.58}$\\
Redshift ($10^{-3}$)  & 0.94 $\pm$ 0.47 \\
Velocity Broadening (\kms)  & 1735$^{+124}_{-109}$\\
Normalization	(10$^{-7}$ cm$^{-5}$)   & 3.74 $\pm$ 0.50\\ \hline \hline
%%%%%%%%%%%%%%%%%%%%%%%%%%%%%%%%%%%%%%%%%%%%%%%%%%%%%%%%%
Fitting to Ar \& Ca Bandpass (2.93--4.7~keV)\tablenotemark{a} \\
$kT_{\rm e}$ (keV)    & 2.04$^{+1.23}_{-0.69}$ \\
Ca (10$^5$ solar) & 1.34$^{+0.26}_{-0.23}$ \\
Ionization Timescale($10^{11}$ cm$^{-3}$ s) & 1.47$^{+4.63}_{-0.69}$\\
Redshift ($10^{-3}$)  & 1.50 $^{+0.69}_{-0.77}$ \\
Velocity Broadening (\kms)  & 1764$^{+178}_{-161}$\\
Normalization	(10$^{-7}$ cm$^{-5}$)   & 3.47$^{+0.69}_{-0.55}$\\ \hline \hline
%%%%%%%%%%%%%%%%%%%%%%%%%%%%%%%%%%%%%%%%%%%%%%%%%%%%%%%%%%
%Fitting to IGE (5.12--7.38~keV) \\
%$kT_{\rm e}$ (keV)    & 4.0 (fixed) \\
%Ionization Timescale($10^{11}$ cm$^{-3}$ s) & 1.47$^{+4.63}_{-0.69}$\\
%Redshift ($10^{-3}$)  & 1.50 $^{+0.69}_{-0.77}$ \\
%Velocity Broadening (\kms)  & 1764$^{+178}_{-161}$\\
%Normalization	(10$^{-7}$ cm$^{-5}$)   & 2.95$^{+0.69}_{-0.55}$\\ \hline \hline
\enddata
%\tablenotetext{a}{Emissivity-weighted centroid of overlapping intercombination doublet}
\tablenotetext{a}{As for the Gaussian fitting, there was not enough signal in the Ar \& Ca lines to each bandpass separately. The electron temperature in particular was unconstrained if attempted to fit separately.}
\end{deluxetable}
%$^{+}_{-}$

\begin{deluxetable}{lcc}
\tablewidth{0pt}
\tablecaption{NEI Fitting to IGEs \label{tab:NEI_IGE_Fitting}}
\tablehead{
\colhead{Parameter} & \colhead{Best-Fit Value (90\% CI)} 
}
\startdata
$N_{\rm{H, MW}}$ (10$^{22}$ cm$^{-2}$)  & 0.062 (fixed) \\
$N_{\rm{H, LMC}}$ (10$^{22}$ cm$^{-2}$)  & 0.3 (fixed) \\ \hline \hline
Swept-Up CSM/ISM: \texttt{nei} \\
kT$_e$ (keV) & 0.546 (fixed) \\
Ionization Timescale ($10^{13}$ cm$^{-3}$ s)  & 5 (fixed) \\
Redshift ($10^{-3}$)  & 1.5 (fixed) \\
Velocity Broadening (\kms)  & 719 (fixed)\\
Normalization	($10^{-2}$ cm$^{-2}$)   & 1.82 (fixed) \\ \hline
%%%%%%%%%%%%%%%%%%%%%%%%%%%%%%%%%%%%%%%%%%%%%%%%%%%%%%%%%%%
Lower-temperature Ejecta Component: \texttt{bvnei} \\
$kT_{\rm e}$ (keV)    & 2.21 (fixed) \\
Abundance (10${^5}$ solar) O--Mg & 0.3 (fixed) \\
%Ne  & 1 (fixed) \\
%Mg  & 3 (fixed)  \\
Si  & 1 (fixed) \\
S  & 1.09 (fixed)\\
Ar  & 1.06 (fixed)\\
Ca  & 1.54 (fixed)\\
Fe  & 0.37 (fixed) \\
Ionization Timescale($10^{11}$ cm$^{-3}$ s) & 1.03$^{+0.42}_{-0.32}$\\
Redshift ($10^{-3}$)  & 2.40$^{+0.37}_{-0.33}$ \\
Velocity Broadening (\kms)  & 1799$^{+200}_{-132}$\\
Normalization	(10$^{-7}$ cm$^{-5}$)   & 3.73$^{+0.86}_{-0.78}$ \\ \hline \hline
%%%%%%%%%%%%%%%%%%%%%%%%%%%%%%%%%%%%%%%%%%%%%%%%%%%%%%%%%
% Below averaged values from my two xwin1 (4-7.4keV, kT2=1.97 keV) and xwin6 (2.5-7 keV, kT2=2.65 keV) fits
Moderate Temperature (4~keV) IGE Component: \texttt{bvvnei} \\
$kT_{\rm e}$ (keV)    & 4 (fixed)\\
Abundance (10${^5}$ solar) Cr & 2.54$^{+2.08}_{-0.96}$ \\
Mn & 0.7 (fixed) \\
Fe & 1 (fixed) \\
Ionization Timescale($10^{9}$ cm$^{-3}$ s) & 7.71$^{+2.59}_{-1.75}$\\
Redshift ($10^{-3}$)  & -3.42$^{+1.43}_{-2.14}$ \\
Velocity Broadening (\kms)  & 2733$^{+676}_{-549}$\\
Normalization	(10$^{-7}$ cm$^{-5}$)   & 1.42$^{+0.23}_{-0.11}$ \\ \hline 
%%%%%%%%%%%%%%%%%%%%%%%%%%%%%%%%%%%%%%%%%%%%%%%%%%%%%%%%%
High Temperature (10~keV) Fe-Ly$\alpha$ component: \texttt{bvvnei}  \\
$kT_{\rm e}$ (keV)    & 10 (fixed)\\
Abundance (10${^5}$ solar) Cr & 58.1 $^{+33.5}_{-21.4}$ \\
Mn & 40.4 $^{+26.7}_{-20.6}$ \\
Fe & 1 (fixed) \\
Ionization Timescale($10^{11}$ cm$^{-3}$ s) & 5.50$^{+2.17}_{-1.32}$\\
Redshift ($10^{-3}$)  & 5.19$^{+1.04}_{-2.64}$ \\ %\tablenotemark{a} \\
Velocity Broadening (\kms)  & 2900 (fixed)\\
Normalization	(10$^{-9}$ cm$^{-9}$)   & 6.19$^{+0.91}_{-1.85}$ \\ \hline 
\enddata
\tablenotetext{}{Results from our spectral fit to the 3.5--7.38~keV bandpass, where we froze all parameters that contributed flux solely to lower energies (i.e., absorption column density, swept-up ambient material, and ejecta abundances below Ca). As the lower-temperature IME-dominated component contributed necessary Fe flux, we allowed its ionization timescale, redshift, velocity broadening, and normalization to vary, but froze the Ca and Fe abundance to ensure it would remain consistent with the observed spectral flux at lower energies.}
\end{deluxetable}

\begin{deluxetable*}{lccccl}
\tablewidth{0pt}
\tablecaption{Emission lines present in Gaussian Fitting\label{tab:Gaussian_Values}}
\tablehead{
\colhead{Transition} & \colhead{Common Name} & \colhead{E$_{\rm rest}$ (eV)} & \colhead{Doppler Shift (\kms)\tablenotemark{b}} & \colhead{$\sigma_{\rm E}$ (eV)\tablenotemark{c}} & \colhead{Norm (10$^{-5}$)}
}
\startdata
\cutinhead{Si \& S band (1.75--3.0~keV)}
Si XIII 2$\rightarrow$1 & Si He$\alpha$ (f) & 1839.4 & 391$^{+161}_{-158}$ & 9.31$^{+0.95}_{-0.84}$ & 14.9$^{+5.61}_{-5.16}$\\
Si XIII 5$\rightarrow$1 \& 6$\rightarrow$1 & Si He$\alpha$ (i) & 1854.0\tablenotemark{a} & - & - & $<$7.28\\
Si XIII 7$\rightarrow$1 & Si He$\alpha$ (r) & 1865.0 & - & - & 50.5$^{+7.53}_{-8.00}$\\
Si XIV 3$\rightarrow$1 \& 4$\rightarrow$1 & Si Ly$\alpha_1$ \& Ly$\alpha_2$ & 2005.5\tablenotemark{a} & - & - & 15.6$^{+2.76}_{-2.44}$\\
Si XIII 13$\rightarrow$1 & Si K$\beta$ & 2182.6 & - & - & 5.40$^{+1.14}_{-1.00}$ \\
Si XIII 23$\rightarrow$1 & - & 2294.1 & - & - &  2.00$^{+0.80}_{-0.68}$\\
Si XIII 37$\rightarrow$1 & - & 2346.0 & - & - & 0.70$^{+0.65}_{-0.54}$ \\
Si XIV 7$\rightarrow$1, 6$\rightarrow$1, \& 67$\rightarrow$1 & Si Ly$\beta_1$ \& Ly$\beta_2$   & 2376.3\tablenotemark{a} & - & - & 2.98$^{+0.90}_{-0.77}$ \\
Si XIV 11$\rightarrow$1 \& 12$\rightarrow$1 & - & 2506.3\tablenotemark{a} & - & - & 0.79$^{+0.54}_{-0.47}$ \\
Si XIV 18$\rightarrow$1 \& 19$\rightarrow$1 & - & 2566.4\tablenotemark{a} & - & - & $<$0.34 %0.277$^{+0.344}_{-0.277}$ 
\\ \hline
S XV 2$\rightarrow$1 & S He$\alpha$ (f) &  2430.3    & 453$^{+139}_{-138}$ & 14.3$^{+1.01}_{-0.92}$ & 7.99$^{+1.75}_{-1.48}$ \\
S XV 5$\rightarrow$1 \& 6$\rightarrow$1 & S He$\alpha$ (i)  & 2447.7\tablenotemark{a} & - & - & $<$1.68 \\
S XV 7$\rightarrow$1 & S He$\alpha$ (r)  & 2460.6  & - & - & 20.4$^{+1.82}_{-1.78}$\\
S XVI 3$\rightarrow$1 \& 4$\rightarrow$1 & S Ly$\alpha_1$ \& Ly$\alpha_2$ & 2622.1\tablenotemark{a} & - & - & 5.30$^{+6.95}_{-6.27}$ \\
S XV 13$\rightarrow$1 & S K$\beta$ & 2884.0 & - & - & 2.84$^{+4.35}_{-3.94}$ \\ %\hline \hline
%\cutinhead{ISM \texttt{vapec}\tablenotemark{a} component} 
%%%%%%%%%%%%%%%%%%%%%%%%%%%%%%%%%%%%%%%%%%%%%%%
%\hline \hline
\cutinhead{Ar \& Ca band (2.75--4.5~keV)}
%Ar \& Ca band (2.75--4.5~keV) \\ \hline
%\sidehead{Emission lines detected in the Ar}
%\cutinhead{Emission lines detected in the Ar \& Ca band (2.75--4.5~keV)} \\ \hline
%& Emission lines detected \\ hline
S XV 13$\rightarrow$1 & S K$\beta$ & 2884.0 & ==453 & ==16.9 & 2.74$^{+0.41}_{-0.38}$ \\
S XV 23$\rightarrow$1 & - & 3032.7 & & & 1.06$^{+0.26}_{-0.24}$ \\
S XVI 6$\rightarrow$1 \& 7$\rightarrow$1 & S Ly$\beta_1$ \& Ly$\beta_2$ & 3106.4\tablenotemark{a} & & & $<$2.28 \\
S XVI 11$\rightarrow$1 \& 12$\rightarrow$1 & - & 3276.2\tablenotemark{a} & & & $<$0.30 \\ \hline
%%%%%%%%%%%%%%%%%%%%%%%%%%%%%%%%%%%%%%%%%%%%%%%
Ar XVII 2$\rightarrow$1 & Ar He$\alpha$ (f) & 3104.1 & 565$^{+559}_{-492}$ & 21.2$^{+4.29}_{-3.31}$ &  $<$2.91\\
Ar XVII 5$\rightarrow$1 \& 6$\rightarrow$1 & Ar He$\alpha$ (i) & 3124.7\tablenotemark{a} & & & $<$4.24\\
Ar XVII 7$\rightarrow$1 & Ar He$\alpha$ (r) & 3139.6 & & & 4.04$^{+1.59}_{-1.92}$ \\
Ar XVII 3$\rightarrow$1 \& 4$\rightarrow$1 & Ar Ly$\alpha_1$ \& Ly$\alpha_2$ & 3321.4 & & & 0.45$^{+0.20}_{-0.18}$\\
ArXVI 10077$\rightarrow$2 \& 10078$\rightarrow$3 & - & 3617 & & & $<$0.09 \\
Ar XVII 13$\rightarrow$1 & Ar K$\beta$ & 3684.5 & & & 0.35\\
Ar XVII 32$\rightarrow$1 & - & 3874.5 & & & $<$1.05 \\
Ar XVII 7$\rightarrow$1 \& 6$\rightarrow$1 & Ar Ly$\beta_1$ \& Ly$\beta_2$ & 3935.2\tablenotemark{a} & & & $<$0.60\\
Ar XVII 37$\rightarrow$1 & - & 3963.7 & & & $<$0.28\\
Ar XVII 55$\rightarrow$1 & - & 4008.7 & & & 0.36$^{+0.14}_{-0.13}$\\ \hline
%%%%%%%%%%%%%%%%%%%%%%%%%%%%%%%%%%%%%%%%%%%%%%%
Ca XIX 2$\rightarrow$1 \& XVIII 10248$\rightarrow$3 & Ca He$\alpha$ (f) & 3861.4\tablenotemark{a} & 176.1$^{+516}_{-491}$ & 17.6$^{+4.12}_{-2.88}$ & 1.15$^{+0.69}_{-0.57}$\\
Ca XVIII 30$\rightarrow$1 & - & 3.8746 & & & $<$2.88 \\
Ca XIX 5$\rightarrow$1 \& 6$\rightarrow$1 & Ca He$\alpha$ (i) & 3.8853\tablenotemark{a} & & & $<$0.80\\
Ca XIX 7$\rightarrow$1 &Ca He$\alpha$ (r)  & 3902.4 & & & 1.06$^{+0.29}_{-0.62}$\\
Ca XX 3$\rightarrow$1 \& 4$\rightarrow$1 & Ca Ly$\beta_1$ \& y$\beta_2$& 4105.0 & & & $<$0.15\\
Ca XIX 13$\rightarrow$1 & Ca K$\beta$ & 4583.5 & & & 0.15$^{+0.08}_{-0.07}$\\
\enddata
\tablenotetext{a}{Emissivity-weighted centroid of overlapping lines}
%\tablenotetext{b}{Emissivity-weighted centroid of overlapping lines}
\tablenotetext{b}{The Doppler shifts off all lines from the same element are equal.}
\tablenotetext{c}{The line widths for all lines from the same element are linked, scaling linearly with energy. }
%\tablenotetext{d}{Not found in XSLIDE, but found in AtomDB web spectrum}
%\tablecomments{\footnotesize{Rest energies are taken from AtomDB XSLIDE}}%, using a mix of identifiable lines and those with $\gtrsim 5 \times 10^{-19}$ cm$^3$ s$^{-1}$} at 1.72~keV (Si \& S) and $\gtrsim 5 \times 10^{-20}$ cm$^3$ s$^{-1}$ (Ar \& Ca). All uncertainties presented reflect 90\% confidence intervals.}
\end{deluxetable*}

%Hello

%% For this sample we use BibTeX plus aasjournalv7.bst to generate the

\bibliography{N103B_XRISM}{}
\bibliographystyle{aasjournalv7}



\end{document}